# Study of Bose-Einstein Condensation Process and Superconductivity

He Wenchen

*School of Science, Hebei University of Technology, 300401, Tianjin, China.*

## Abstract

The transition from normal state to superconductive state is a condensation (Bose-Einstein condensation) process. As an example of Bose-Einstein condensation, the condensation process of superconductivity is investigated in detail. The basic unit of superconductivity is condensation circuit, which should contain 4m+2 (m=1,2,...) electrons (holes). The perfect explanation for no resistance and complete diamagnetism are given by using condensation circuit. The quantum mechanics equation of superconductive state is constructed, and energy level equation of superconductive quantum state is given. The later equation has simple form $x\boldsymbol{I} - \boldsymbol{A} = 0$ ($I$ is the unit matrix, $A$ is the adjacent matrix), with deep topological meanings. The phase transition from normal state to superconductive state is irreversible. The irreversible process is analyzed and discussed in detail. The temperature region $0 < T < T_c$ of superconductive state is calculated. The simplest structure of superconductor circuit is a condensation circuit which is composed of a series of 3-4-element-circuits (called lantern series structure). It is theoretically proved that the room temperature superconductivity can be realized by using polylayer oxide structure. The experimental frame of room temperature superconductivity is suggested.

## I. Introduction

Since 1911 Onnes discovered superconductivity [1] more than 110 years have passed. Many researchers have made beneficial contributions in this field [1-6]. In

Corresponding author at: School of Science, Hebei University of Technology, Tianjin 300401, People's Republic of China. E-mail address: hwch@hebut.edu.cn (He Wenchen)

1986, Bednorz and Műller discovered superconductor with high transition temperature [5]. People's interest start to concentrate on the microscopic mechanism of high temperature superconductors and discovered many superconductors with high transition temperature [7-9]. However, up to now the problems concerned the microscopic mechanism of superconductor especially high temperature superconductor is open. In addition, it is unknown that the room temperature superconductivity can be realized or not. These problems are investigated in detail in this contribution.

# II. Theoretical formalism

Superconductive phase is a low temperature condensation phase. The transition from normal phase to superconductive phase is an irreversible condensation process (Bose-Einstein condensation). This condensation process should include the following two aspects:

I). Condensation circuits (electron or hole circuits) generate.

II). Condensation circuits contract and tend to have the minimum area.

These circuits are the basic unit of the superconductor. In the whole condensation process, the electrostatic balance and central symmetry are remained all the time. For every aspect of condensation process, the electron distribution graphs are obtained and its generation energy is calculated.

## A. Quantum mechanics equation of an electron cluster.

To begin with, I calculate the generation energy for any electron cluster containing $n$ electrons, and the Hamiltonian equation of electron cluster in a crystal lattice is as follows.

$$\mathcal{H} = -\frac{\hbar^2}{2m}\sum_i^n \nabla_i^2 + \sum_i^n U_{lattice,i} + \frac{1}{2}\sum_i^n \sum_{j\neq i}^n \left(\frac{e^2}{r_{ij}}\right), \tag{1}$$

in which the first term is the $n$ electrons' kinetic energy, and the second term is the interaction energy between the $n$ electrons and all the lattice points (atom kernels), and the third one is the interaction energy among the $n$ electrons in the cluster. Formula (1) can be simply written as follows,

$$\mathcal{H}^{eff} = \sum_{i=1}^{n} \mathcal{H}_i^{eff} \tag{2}$$

$$\mathcal{H}_i^{eff} = -\frac{\hbar^2}{2m}\nabla_i^2 + U_{lattice,i} + \frac{n-1}{2}\left(\frac{e^2}{r_{ij}}\right)_{average\ over\ j}, \tag{3}$$

in which $\mathcal{H}_i^{eff}$ is the effective Hamiltonian operator for single electron.

Corresponding author at: School of Science, Hebei University of Technology, Tianjin 300401, People's Republic of China. E-mail address: hwch@hebut.edu.cn (He Wenchen)

Considering (2) and (3), we have that

$$\mathcal{H}_i^{eff}\psi_i(i)=\varepsilon_i\psi_i(i)\quad(i=1,2,\cdots,n)\qquad(4)$$

$$\mathcal{H}^{eff}\prod_i^n\psi_i(i)=\varepsilon\prod_i^n\psi_i(i)\ ,\qquad(5)$$

$$\mathcal{H}^{eff}\psi=\varepsilon\psi\qquad(6)$$

in which

$$\varepsilon=\sum_i^n\varepsilon_i\ ,\quad \mathcal{H}^{eff}=\sum_i^n\mathcal{H}_i^{eff}\ ,\quad \psi=\prod_i^n\psi_i\quad(i=1,2,\cdots,n)\qquad(7)$$

$$\psi_i=\sum_{r=1}^n c_{ir}\varphi_r.\qquad(8)$$

Substituting (8) into (5), we obtain that

$$\sum_{r=1}^n c_{ir}[\mathcal{H}^{eff}-\varepsilon]\varphi_r=0\quad(i=1,2,\cdots,n)\qquad(9)$$

Left time $\varphi_s$* to formula (9) and integrate over the whole space. We have that

$$\sum_{s=1}^n c_{ir}[H_{rs}^{eff}-\varepsilon S_{rs}]=0\quad(i=1,2,\cdots,n)\qquad(10)$$

in which $S_{rs}$, $H_{rs}^{eff}$ is the overlap integral and the energy integral, respectively.

$$\begin{cases}H_{rs}=H_{sr}=\int\varphi_r^*\mathcal{H}^{eff}\varphi_s d\tau=\begin{cases}\beta_{rs} & (r \text{ is adjacent to } s)\\ \alpha_r & (r=s)\\ 0 & (\text{the other cases})\end{cases}\\ S_{rs}=S_{sr}=\int\varphi_r^*\varphi_s d\tau=\delta_{rs}\end{cases}\qquad(11)$$

Formula (10) can be written as (12)

$$\begin{pmatrix}\alpha_1-\varepsilon & \beta_{12} & \cdots & \beta_{1n}\\ \beta_{21} & \alpha_2-\varepsilon & \cdots & \beta_{2n}\\ \vdots & \vdots & \vdots & \vdots\\ \beta_{n1} & \beta_{n2} & \cdots & \alpha_n-\varepsilon\end{pmatrix}\begin{pmatrix}c_{i1}\\ c_{i2}\\ \vdots\\ c_{in}\end{pmatrix}=0\quad(i=1,2,...,n)\qquad(12)$$

The secular equation of (12) can be written as (13)

$$\begin{vmatrix}\alpha_1-\varepsilon & \beta_{12} & \cdots & \beta_{1n}\\ \beta_{21} & \alpha_2-\varepsilon & \cdots & \beta_{2n}\\ \vdots & \vdots & \vdots & \vdots\\ \beta_{n1} & \beta_{n2} & \cdots & \alpha_n-\varepsilon\end{vmatrix}=0\ .\quad(i=1,2,...,n)\qquad(13)$$

Its solutions $\varepsilon_1$, $\varepsilon_2,...,\varepsilon_n$ (suppose that $\varepsilon_1\le\varepsilon_2\le...\le\varepsilon_n$ ) are the eigen energies of $H^{eff}$.

If the cluster is a closed circuit, then the formula (11) becomes (14), and the formula (12 ) becomes (15).

$$H_{rs}=\begin{cases}\beta & (r \text{ is adjacent to s})\\ \alpha_r=\alpha & (r=s)\\ 0 & (\text{the other cases})\end{cases}\qquad(14)$$

$$\begin{pmatrix}\alpha-\varepsilon & \beta & 0 & \cdots & \beta\\ \beta & \alpha-\varepsilon & \beta & 0 & \cdots\\ \cdots & \ddots & \ddots & \ddots & \cdots\\ \cdots & \cdots & \beta & \alpha-\varepsilon & \beta\\ \beta & 0 & 0 & \beta & \alpha-\varepsilon\end{pmatrix}\begin{pmatrix}c_{i1}\\ c_{i2}\\ \cdots\\ \cdots\\ c_{in}\end{pmatrix}=0\qquad(15)$$

The secular equation of (15) can be written as (16).

$$\begin{vmatrix}\alpha-\varepsilon & \beta & 0 & \cdots & \beta\\ \beta & \alpha-\varepsilon & \beta & 0 & \cdots\\ \cdots & \ddots & \ddots & \ddots & \cdots\\ \cdots & \cdots & \beta & \alpha-\varepsilon & \beta\\ \beta & 0 & 0 & \beta & \alpha-\varepsilon\end{vmatrix}=0\qquad(16)$$

Let

$$x_i=\frac{\varepsilon_i-\alpha}{-\beta}\qquad(17)$$

$$E_g=2\sum_{i=1}^{n/2}\varepsilon_i=2\sum_{i=1}^{n/2}(-\beta x_i+\alpha)\quad(n=even)\qquad(18)$$

Where the factor 2 comes from that there are two electrons in one energy level. $-2E_g$ is the generation energy of the ground state of the electron cluster system. It describes the stability of the cluster.

The secular equation (16) can be simplified into（19）

$$\begin{vmatrix}x & -1 & 0 & \cdots & -1\\ -1 & x & -1 & 0 & \cdots\\ 0 & \ddots & \ddots & \ddots & 0\\ \cdots & 0 & -1 & x & -1\\ -1 & \cdots & 0 & -1 & x\end{vmatrix}_{n\times n}=0\ .\qquad(19)$$

In fact, $x$ is nothing else but the eigen energy of $H^{eff}$. It is the eigen energy relative to $\alpha$ and in the unit $-\beta$ $(\beta<0)$. The formula (19) can be written as (20).

$$x\boldsymbol{I}-\boldsymbol{A}=0,\qquad(20)$$

in which $\boldsymbol{I}$(/$\boldsymbol{A}$) is the unit (/electron adjacent) matrix. Equation (20) clearly shows that the

Corresponding author at: School of Science, Hebei University of Technology, Tianjin 300401, People's Republic of China. E-mail address: hwch@hebut.edu.cn (He Wenchen)

topological theory and the quantum theory have closed relation. Obviously, the physics and chemistry properties of the cluster depend largely on the topological properties [10]. From (19), we can obtain that

$$\prod_{p=1}^{n}(x-e^{i\frac{2p\pi}{n}}-e^{-i\frac{2p\pi}{n}})=\prod_{p=-n/2}^{n/2-1}(x-2\cos\frac{2p\pi}{n})=0 \tag{21}$$

$$x_{closed}=2\cos\frac{2p\pi}{n},\qquad (p=0,\pm1,\pm2,\cdots,\pm(\frac{n}{2}-1),+\frac{n}{2}) \tag{22}$$

As comparison, I calculate the energy levels of an opened $n$-electron circuits. Follow the similar steps as those from (17) to (19). For this opened circuit system I obtain that

$$\begin{vmatrix} x & -1 & 0 & \cdots & 0 \\ -1 & x & -1 & 0 & \cdots \\ 0 & \ddots & \ddots & \ddots & 0 \\ \cdots & 0 & -1 & x & -1 \\ 0 & \cdots & 0 & -1 & x \end{vmatrix}_{n\times n}=0 \tag{23}$$

$$\prod_{p=1}^{n}(x-2\cos\frac{p\pi}{n+1})=0 \tag{24}$$

$$x_{opened}=2\cos\frac{p\pi}{n+1}\qquad (p=1,2,\cdots,\ n) \tag{25}$$

Definition: The cyclization energy $E_c$ is the ground states energy difference between the opened circuit and the closed circuit.

$$E_c=E_{opened}-E_{closed} \tag{26}$$

When $n=4m$, we have that

$$E_{closed,n=4m}=4m\alpha+2\beta\left[1+2\sum_{j=1}^{m-1}\cos\frac{2j\pi}{4m}-1\right] \tag{27}$$

$$E_{opened,n=4m}=4m\alpha+2\beta\left[\sum_{j=1}^{m}\cos\frac{(2j-1)\pi}{4m+1}+\sum_{j=1}^{m}\cos\frac{2j\pi}{4m+1}\right] \tag{28}$$

$$\begin{aligned} E_{c,n=4m} &= 2\beta\left[\sum_{j=0}^{m-1}\cos\frac{(2j+1)\pi}{4m+1}+\sum_{j=0}^{m-1}\cos\frac{2(j+1)\pi}{4m+1}\right]-2\beta\left[\sum_{j=0}^{m-1}\cos\frac{2(j+1)\pi}{4m}+\sum_{j=0}^{m-1}\cos\frac{2j\pi}{4m}\right] \\ &=4\beta\left[\sum_{j=0}^{m-1}\cos\frac{(2j+1)\pi}{4m+1}\cos\frac{\pi}{4m+1}\right]-4\beta\left[\sum_{j=0}^{m-1}\cos\frac{(2j+1)\pi}{4m}\cos\frac{\pi}{4m}\right]<0 \\ &\left(\cos\frac{(2j+1)\pi}{4m+1}>\cos\frac{(2j+1)\pi}{4m},\quad \cos\frac{\pi}{4m+1}>\cos\frac{\pi}{4m}\right) \end{aligned} \tag{29}$$

*When* $n=4m+2$, we have that

$$E_{opened,n=4m+2}=(4m+2)\alpha+2\beta\left[\sum_{j=0}^{m}\cos\frac{2(j+1)\pi}{4m+3}+\sum_{j=1}^{m}\cos\frac{2j\pi}{4m+3}\right] \tag{30}$$

$$\begin{aligned} E_{closed,n=4m+2} &=(4m+2)\alpha+2\beta\left[\sum_{j=0}^{m}\cos\frac{2j\pi}{4m+2}+\sum_{j=1}^{m}\cos\frac{-2j\pi}{4m+2}\right] \\ &=(4m+2)\alpha+2\beta+4\beta\sum_{j=1}^{m}\cos\frac{2j\pi}{4m+2} \end{aligned} \tag{31}$$

$$\begin{aligned} E_{c,n=4m+2} &=2\beta\left[\sum_{j=0}^{m}\cos\frac{2(j+1)\pi}{4m+3}+\sum_{j=1}^{m}\cos\frac{2j\pi}{4m+3}\right]-2\beta\left[\sum_{j=0}^{m}\cos\frac{2j\pi}{4m+2}+\sum_{j=1}^{m}\cos\frac{-2j\pi}{4m+2}\right] \\ &=2\beta\left(\cos\frac{\pi}{4m+3}-1\right)-4\beta\left[\sum_{j=1}^{m}\cos\frac{2j\pi}{4m+2}-\sum_{j=1}^{m}\cos\frac{(2j+1)\pi}{4m+3}\cos\frac{\pi}{4m+3}\right]>0 \\ &(\cos\frac{(2j+1)\pi}{4m+3}<\cos\frac{2j\pi}{4m+2};\quad \cos\frac{\pi}{4m+3}<1) \end{aligned} \tag{32}$$

(29) and (32) show that for $n$=4m, $E_c<0$ ; and for $n$=4m+2, $E_c>0$. The 4m+2 circuit can be called the condensation circuit stated above.

## B. The illustration of condensation process

Stage 1. At the temperature lower than critical temperature $T_C$, (4m+2) electron circuit (m=1,2,...) will close to generate condensation circuits, shown in Fig.1.

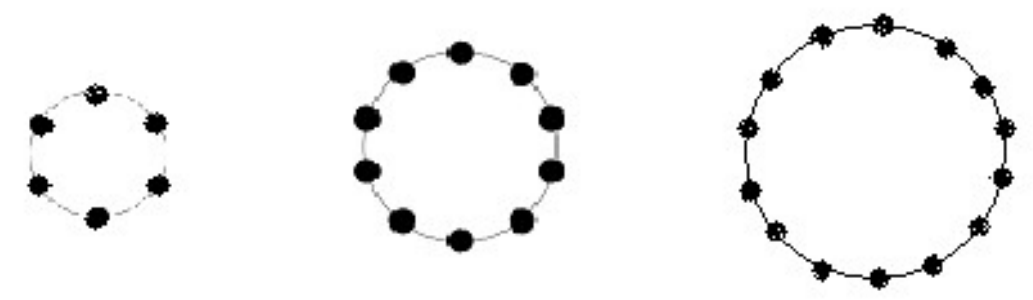

m=1 m=2 m=3

FIG. 1. The condensation circuit for m=1, 2, 3, respectively.

Stage 2. The condensation circuits will contract and tend to have the minimum area. For m=1, 2, 3...(or $n$=6,10,14), the 4m+2 charges (grey round point) at the corners or intersection points in the cluster will move to the middle of the 4m+2 edges and further reach electrostatic equilibrium.

Corresponding author at: School of Science, Hebei University of Technology, Tianjin 300401, People's Republic of China. E-mail address: hwch@hebut.edu.cn (He Wenchen)

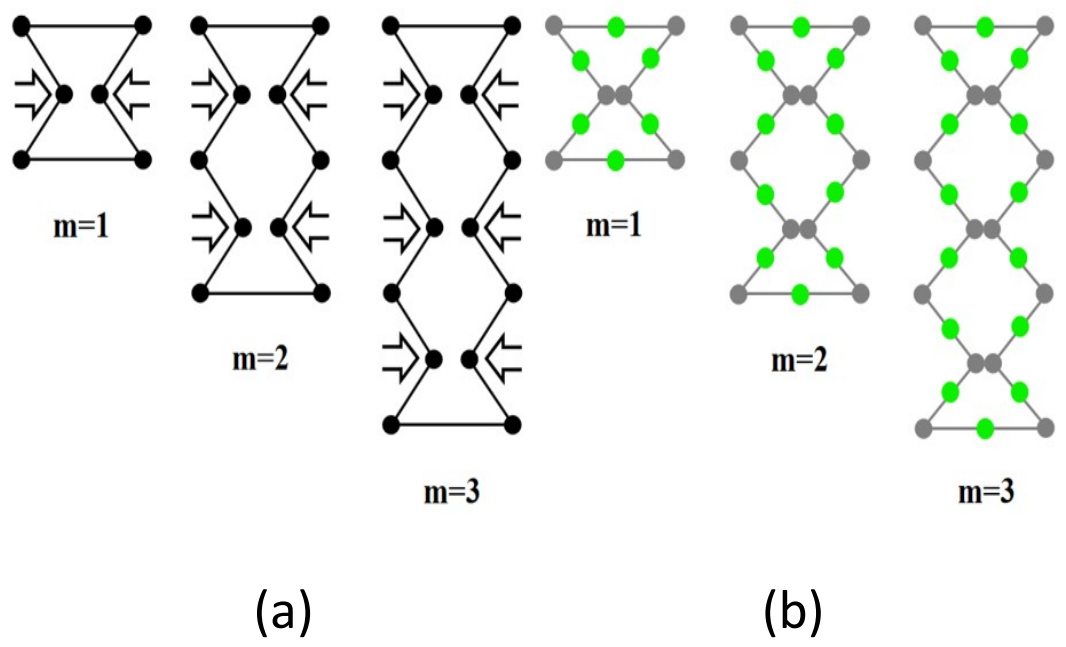

FIG. 2. (a) condensation process (b) final case and charge redistribution (from corners and intersections to the middle of edge).

Fig.2(a) shows that the condensation circuits tend to have minimum area. Fig.2(b) shows that for electrostatic balance the charges (electrons or holes) at the corners or intersections points of the cluster will move to the middle of the 4m+2 edges, as a result forming a series of 3-4 element circuit, i.e. the minimum superconductivity unit (called lantern series structure) shown in Fig.3, where *m* is the number of intersections. This superconductivity unit include two triangles (at the top and bottom) and m squares, of which the center has an intersection.

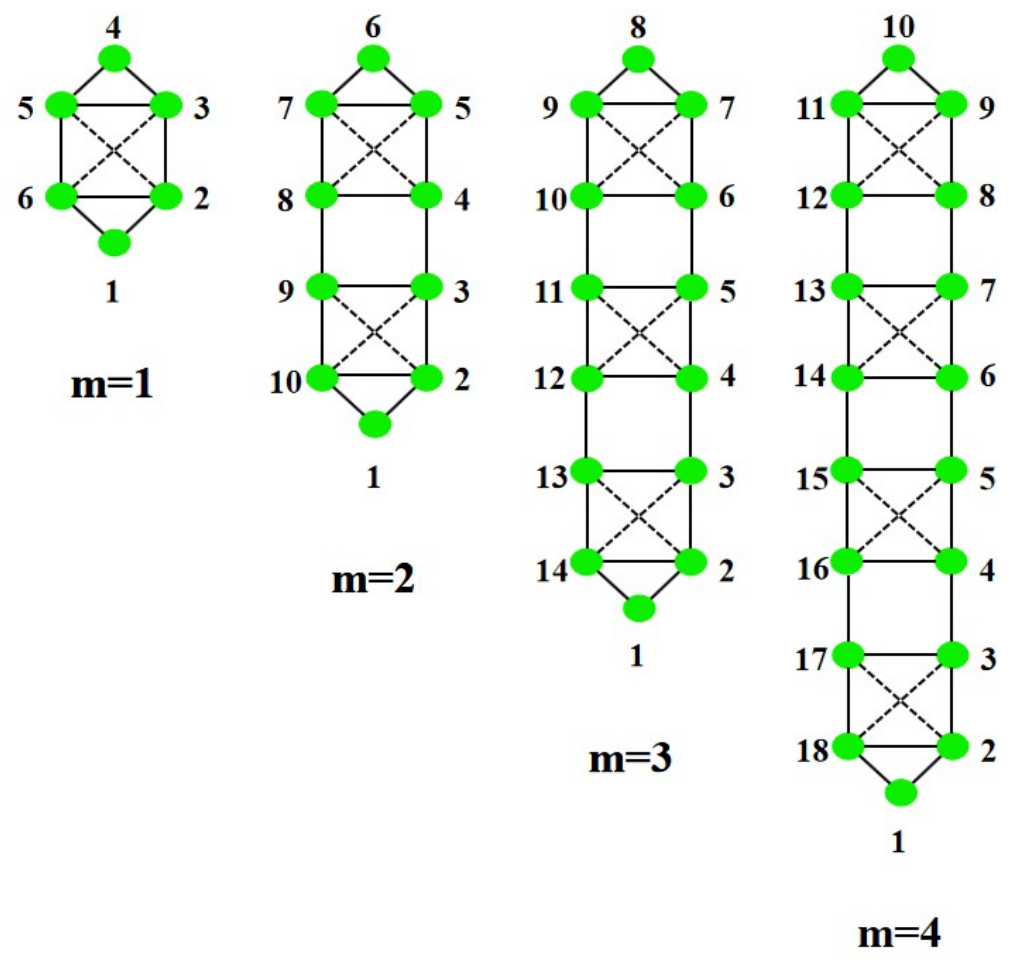

FIG. 3. Minimum superconductivity unit for m=1, 2, 3, 4, respectively.

## C. Electron energy levels and generation energy of minimum superconductive unit

According to formula (20) we write down the following secular equation of Hamiltonian for minimum superconductivity unit in Fig.3 for m=1, 2, 3, 4, respectively.

$$\begin{vmatrix} x & -1 & 0 & 0 & 0 & -1 \\ -1 & x & -1 & 0 & -1 & -1 \\ 0 & -1 & x & -1 & -1 & -1 \\ 0 & 0 & -1 & x & -1 & 0 \\ 0 & -1 & -1 & -1 & x & -1 \\ -1 & -1 & -1 & 0 & -1 & x \end{vmatrix}_{6\times 6} =0, \ (m=1)$$

$$\begin{vmatrix} x & -1 & 0 & 0 & 0 & 0 & 0 & 0 & 0 & -1 \\ -1 & x & -1 & 0 & 0 & 0 & 0 & 0 & -1 & -1 \\ 0 & -1 & x & -1 & 0 & 0 & 0 & 0 & -1 & -1 \\ 0 & 0 & -1 & x & -1 & 0 & -1 & -1 & 0 & 0 \\ 0 & 0 & 0 & -1 & x & -1 & -1 & -1 & 0 & 0 \\ 0 & 0 & 0 & 0 & -1 & x & -1 & 0 & 0 & 0 \\ 0 & 0 & 0 & -1 & -1 & -1 & x & -1 & 0 & 0 \\ 0 & 0 & 0 & -1 & -1 & 0 & -1 & x & -1 & 0 \\ 0 & -1 & -1 & 0 & 0 & 0 & 0 & -1 & x & -1 \\ -1 & -1 & -1 & 0 & 0 & 0 & 0 & 0 & -1 & x \end{vmatrix}_{10\times 10} =0, (m = 2)$$

$$\begin{vmatrix} x & -1 & 0 & 0 & 0 & 0 & 0 & 0 & 0 & 0 & 0 & 0 & 0 & -1 \\ -1 & x & -1 & 0 & 0 & 0 & 0 & 0 & 0 & 0 & 0 & 0 & -1 & -1 \\ 0 & -1 & x & -1 & 0 & 0 & 0 & 0 & 0 & 0 & 0 & 0 & -1 & -1 \\ 0 & 0 & -1 & x & -1 & 0 & 0 & 0 & 0 & 0 & -1 & -1 & 0 & 0 \\ 0 & 0 & 0 & -1 & x & -1 & 0 & 0 & 0 & 0 & -1 & -1 & 0 & 0 \\ 0 & 0 & 0 & 0 & -1 & x & -1 & 0 & -1 & -1 & 0 & 0 & 0 & 0 \\ 0 & 0 & 9 & 0 & 0 & -1 & x & -1 & -1 & -1 & 0 & 0 & 0 & 0 \\ 0 & 0 & 0 & 0 & 0 & 0 & -1 & x & -1 & 0 & 0 & 0 & 0 & 0 \\ 0 & 0 & 0 & 0 & 0 & -1 & -1 & -1 & x & -1 & 0 & 0 & 0 & 0 \\ 0 & 0 & 0 & 0 & 0 & -1 & -1 & 0 & -1 & x & -1 & 0 & 0 & 0 \\ 0 & 0 & 0 & -1 & -1 & 0 & 0 & 0 & 0 & -1 & x & -1 & 0 & 0 \\ 0 & 0 & 0 & -1 & -1 & 0 & 0 & 0 & 0 & 0 & -1 & x & -1 & 0 \\ 0 & -1 & -1 & 0 & 0 & 0 & 0 & 0 & 0 & 0 & 0 & -1 & x & -1 \\ -1 & -1 & -1 & 0 & 0 & 0 & 0 & 0 & 0 & 0 & 0 & 0 & -1 & x \end{vmatrix}_{14\times 14} =0, \quad (m = 3)$$

Corresponding author at: School of Science, Hebei University of Technology, Tianjin 300401, People's Republic of China. E-mail address: hwch@hebut.edu.cn (He Wenchen)

$$\begin{vmatrix}
x & -1 & 0 & 0 & 0 & 0 & 0 & 0 & 0 & 0 & 0 & 0 & 0 & 0 & 0 & 0 & 0 & -1 \\
-1 & x & -1 & 0 & 0 & 0 & 0 & 0 & 0 & 0 & 0 & 0 & 0 & 0 & 0 & 0 & -1 & -1 \\
0 & -1 & x & -1 & 0 & 0 & 0 & 0 & 0 & 0 & 0 & 0 & 0 & 0 & 0 & 0 & -1 & -1 \\
0 & 0 & -1 & x & -1 & 0 & 0 & 0 & 0 & 0 & 0 & 0 & 0 & 0 & -1 & -1 & 0 & 0 \\
0 & 0 & 0 & -1 & x & -1 & 0 & 0 & 0 & 0 & 0 & 0 & 0 & 0 & -1 & -1 & 0 & 0 \\
0 & 0 & 0 & 0 & -1 & x & -1 & 0 & 0 & 0 & 0 & 0 & -1 & -1 & 0 & 0 & 0 & 0 \\
0 & 0 & 0 & 0 & 0 & -1 & x & -1 & 0 & 0 & 0 & 0 & \bar{1} & -1 & 0 & 0 & 0 & 0 \\
0 & 0 & 0 & 0 & 0 & 0 & -1 & x & -1 & 0 & -1 & -1 & 0 & 0 & 0 & 0 & 0 & 0 \\
0 & 0 & 0 & 0 & 0 & 0 & 0 & -1 & x & -1 & -1 & -1 & 0 & 0 & 0 & 0 & 0 & 0 \\
0 & 0 & 0 & 0 & 0 & 0 & 0 & 0 & -1 & x & -1 & 0 & 0 & 0 & 0 & 0 & 0 & 0 \\
0 & 0 & 0 & 0 & 0 & 0 & 0 & -1 & -1 & -1 & x & -1 & 0 & 0 & 0 & 0 & 0 & 0 \\
0 & 0 & 0 & 0 & 0 & 0 & 0 & -1 & -1 & 0 & -1 & x & -1 & 0 & 0 & 0 & 0 & 0 \\
0 & 0 & 0 & 0 & 0 & -1 & -1 & 0 & 0 & 0 & 0 & -1 & x & -1 & 0 & 0 & 0 & 0 \\
0 & 0 & 0 & 0 & 0 & -1 & -1 & 0 & 0 & 0 & 0 & 0 & -1 & x & -1 & 0 & 0 & 0 \\
0 & 0 & 0 & -1 & -1 & 0 & 0 & 0 & 0 & 0 & 0 & 0 & 0 & -1 & x & -1 & 0 & 0 \\
0 & 0 & 0 & -1 & -1 & 0 & 0 & 0 & 0 & 0 & 0 & 0 & 0 & 0 & -1 & x & -1 & 0 \\
0 & -1 & -1 & 0 & 0 & 0 & 0 & 0 & 0 & 0 & 0 & 0 & 0 & 0 & 0 & -1 & x & -1 \\
-1 & -1 & -1 & 0 & 0 & 0 & 0 & 0 & 0 & 0 & 0 & 0 & 0 & 0 & 0 & 0 & -1 & x
\end{vmatrix}_{18\times 18} = 0, \quad (m = 4)$$

The solutions of the above four secular equations are as follows,

For m=1, x=-2, -1, -1; -0.5616, 1, 3.5616.

For m=2, x= -2, -2, -1.4893, -1, -1; 0, 0, 0.7108, 3, 3.7785.

For m=3, x=-2, -2, -2, -1.74,-1, -1, -1; 0, 0, 0.225, 0.5858, 2.65,3.4142, 3.86.

For m=4, x=-2,-2,-2, -1.81, -1.81， -1.35, -1,-1， -0.72; 0, 0, 0.326, 0.326， 0.54， 2.12, 3.036， 3.53, 3.8.

All the energy levels are composed of upper half part energy band and lower half part energy band. The summations of energy levels for upper part and lower part bands are equal and opposite.

We define $E_s$=-β $\sum_{i=1}^{n/2} x_i$ as the absolute value of summations of energy levels for upper (lower) part band. For m=1, 2, 3, 4, $E_s$=-4β, -7.49β, -10.74β, -13.69β, respectively, listed in Table 1. 2$E_s$ is the ground state energy of superconductive state, and the relation between $E_s$ and $E_g$ is as follows.

$$E_S = E_g + n\alpha \tag{33}$$

As for m=0 ($n$=2) the condensation circuit does not exist. For the size of electron (or hole) pair the only constraint is from the uncertainty principle. So the size $l$ of electron (or hole) pair

$$l > \frac{\hbar}{p_F} = \frac{h/2\pi}{p_F} \tag{34}$$

In which $p_F$ is the Fermi energy level and $h$ is Planck constant.

To begin with, we supposed that the $n$ (4m+2) electrons (or holes) had no interaction and the pauli exclusion principle wasn't considered, and they all would occupy the same level. Now, we consider the interaction of the $n$ electrons (or holes) and then the levels will split into $\varepsilon_1, \varepsilon_2, ..., \varepsilon_n$ . One half of levels, $\varepsilon_1, \varepsilon_2, \cdots, \varepsilon_{n/2}$ will decrease, and the other half of , $\varepsilon_{(n/2)+1}, \varepsilon_{(n/2)+2}, \cdots, \varepsilon_n$ will increase. The summation of $\varepsilon_{(n/2)+1}, \varepsilon_{(n/2)+2}, \cdots, \varepsilon_n$ is $E_s$, and the summation of $\varepsilon_1, \varepsilon_2, \cdots, \varepsilon_{n/2}$ is -$E_s$. Now let all the energy levels shift $E_s$/(n/2), and the energy levels for lower half part bands become $\varepsilon_1 + E_s/(n/2), \varepsilon_2 + E_s/(n/2), ..., \varepsilon_{n/2} + E_s/(n/2),$ , while the energy levels for upper half part bands become

Corresponding author at: School of Science, Hebei University of Technology, Tianjin 300401, People's Republic of China. E-mail address: hwch@hebut.edu.cn (He Wenchen)

$\varepsilon_{(n/2)+1} + E_s/(n/2), \varepsilon_{(n/2)+2} + E_s/(n/2), \cdots, \varepsilon_n + E_s/(n/2)$. When the temperature increases, the upper half part bands are occupied gradually by electrons (or holes) and completely occupied till the temperature reaches $T_c$. When $T > T_c$, the electrons will escape out of upper half part bands and enter into the normal state region. Therefore, the upper limit of temperature region for superconductive state is $T_c$.

$$\frac{1}{2}kT_c = 2\sum_{i=n/2}^{n}[\varepsilon_i + E_s/(n/2)]=4E_s \qquad (35)$$

$$T_c = \frac{8E_s}{k} \qquad (36)$$

in which $k$ is the Boltzmann constant. When the temperature decreases, the lower half part bands are occupied gradually by electrons (or holes) and completely occupied till the temperature reaches $T_0$.

$$\frac{1}{2}kT_0 = 2\sum_{i=1}^{n/2}[\varepsilon_i + E_s/(n/2)]=0 \qquad (37)$$

And $T_0 = 0$.

Then the system is called superconductive ground state or Motte insulator. Thus the temperature region of superconductivity is $0<T<T_c$. While, the temperature region of normal state is $T>T_c$.

When $m$=1, $T_c$ =93K which is taken from the experimental result [8], β can be calculated. And $T_c$'s for $m$=0,1,2,3,4 are shown in table 1. This table indicates that if $m\geq4$, then $T_c\geq314$K. Hence, we have the conclusion that the preparation of the room temperature superconductor is possible.

Table 1. The $E_s$ and critical temperature $T_c$

| | m=0 | m=1 | m=2 | m=3 | m=4 |
|---|---|---|---|---|---|
| $E_s$ | -β | -4β | -7.49β | -10.74β | -13.69β |
| $T_c$ | 23K | 93K[7] | 167K | 243K | 314K |

## D. How to manufacture the room temperature superconductor (experimental frame)

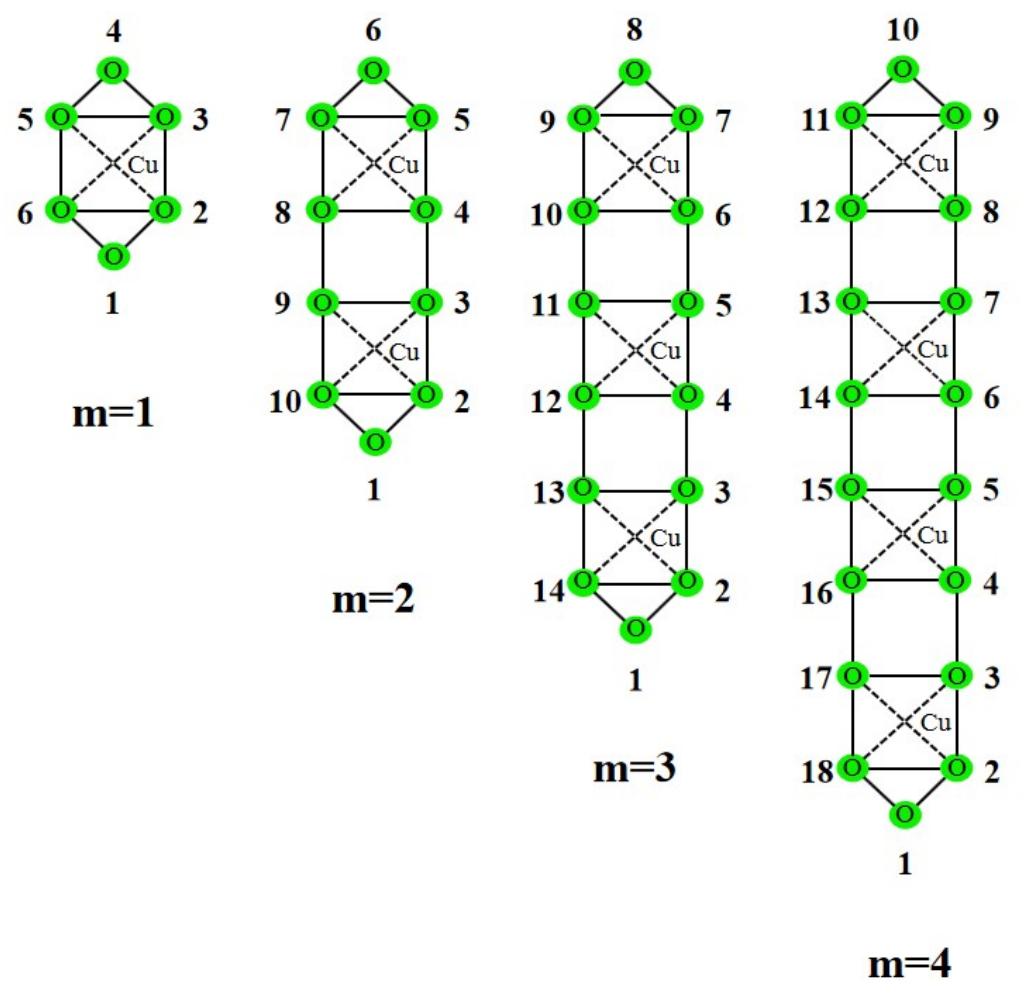

Corresponding author at: School of Science, Hebei University of Technology, Tianjin 300401, People's Republic of China. E-mail address: hwch@hebut.edu.cn (He Wenchen)

FIG. 4. Crystal cell of oxide copper type for m=1, 2, 3, 4, respectively.

Crystal cell structure of oxide copper type is shown in Fig.4. It is similar to Fig.3, except that the intersect points and the corner points are replaced by copper and oxygen, respectively. The steps of manufacturing such superconductors are suggested as follows,

**1.** Take the matching ratio of copper and oxygen as m:(2m+1) (the ratio of the mole numbers). If in reactants there exist only copper and oxygen, then the products is oxide copper eigen superconductor in which electron carrier number density is equal to hole carrier number density).

If add Y and Ba in reactants (Y:Ba:Cu is equal to 1:2:3), we can obtain the doped superconductors (*p* or *n* type).

**2**. Put the matched reactants into furnace to sinter. The sintering temperature is larger than $T_c$.

### E. The transition from normal state to superconductive state is an irreversible process.

When the temperature increases and reaches $T_c$ and the system enters into normal state. When the temperature decreases from the higher temperature than $T_c$ to the lower than $T_c$ the system does not immediately enter into superconductive state because the energy levels of superconductive state have not been constructed in that time. Until the temperature is equal to $(1/2)T_c$, the energy levels of superconductive state begin to be constructed. The transition processes are given in Fig.5. It is found that the two curves do not coincide with each other. This indicates that the phase transition from normal phase to superconductive phase is irreversible.

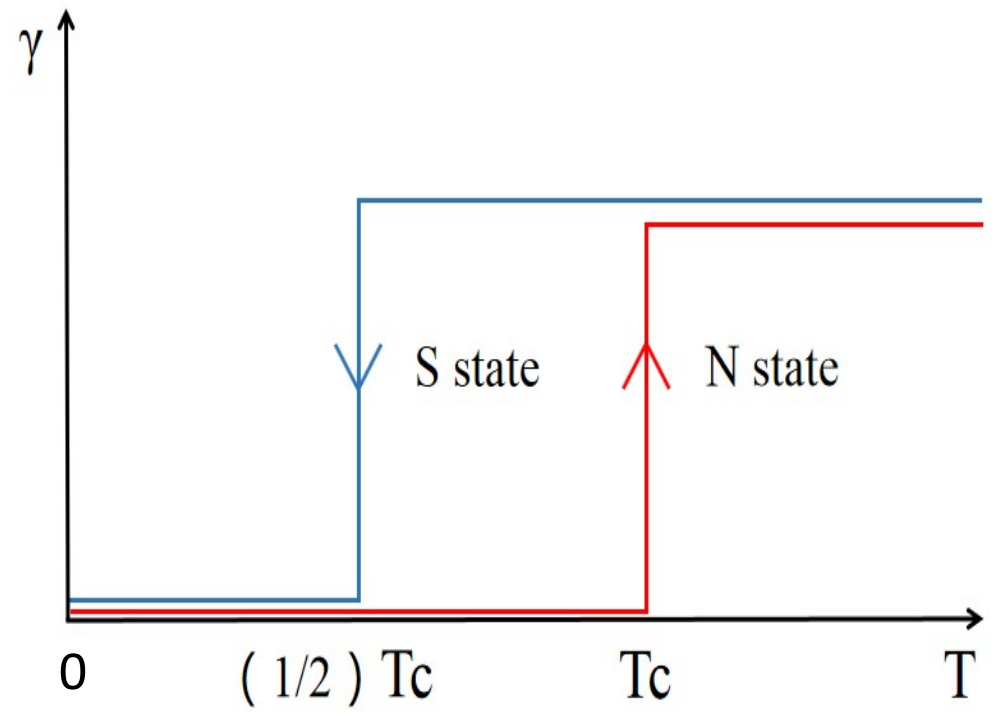

FIG. 5. Superconductive transition process

## III.Conclusions and discussion

1. Superconductive unit is electron (hole) circuit which includes 4m+2 electrons (holes) . Such a circuit is formed by condensation. If take m=0, then 4m+2=2, the circuits degenerate into pairs, and this is the case of BCS' low temperature superconductivity.

2. If the temperature doesn't change, then the electron (hole) distribution on the circuit will not change. The total momentum modulus

Corresponding author at: School of Science, Hebei University of Technology, Tianjin 300401, People's Republic of China. E-mail address: hwch@hebut.edu.cn (He Wenchen)

and angular momentum modulus of each electron (hole) circuit (called condensation circuit) unchange all the time in the superconductive state. All the electrons on the circuit are located on the bond edges. A circuit has no interaction force and no force moment with lattice and other circuits. That is why the superconductor has no resistance.

3. The condensation circuits themselves play a part in diamagnetism. This is why the superconductor has complete diamagnetism.

4. The β reflects the interaction of electrons (holes) on the circuit. The value of β changes with different material, and it can also affect the $T_c$ of superconductor.

Corresponding author at: School of Science, Hebei University of Technology, Tianjin 300401, People's Republic of China. E-mail address: hwch@hebut.edu.cn (He Wenchen)

Corresponding author at: School of Science, Hebei University of Technology, Tianjin 300401, People’s Republic of China. E-mail address: hwch@hebut.edu.cn (He Wenchen)